\documentclass[12pt]{iopart}
\usepackage{latexsym}
\usepackage{epsfig}

\usepackage{iopams}
\usepackage{amsthm}
\usepackage{url}
\usepackage{hyperref}

\newcommand{\ket}[1]{\left | \, #1 \right \rangle}
\newcommand{\bra}[1]{\left \langle #1 \, \right |}
\newcommand{\proj}[2]{\ket{#1}\!\!\bra{#2}}

\newtheorem{theorem}{Theorem}

\newtheorem{lemma}{Lemma}
\newtheorem{corollary}{Corollary}

\begin{document}

\title[DIQKD secure against coherent attacks]{Device independent quantum key distribution secure against coherent attacks with memoryless measurement devices}
\author{Matthew McKague}
\address{University of Waterloo, Waterloo N2L 3G1, Canada}
\eads{memckagu@uwaterloo.ca}

\maketitle
\bibliographystyle{halphads}
\begin{abstract}
Device independent quantum key distribution aims to provide a higher degree of security than traditional QKD schemes by reducing the number of assumptions that need to be made about the physical devices used.  The previous proof of security by Pironio et al. applies only to collective attacks where the state is identical and independent and the measurement devices operate identically for each trial in the protocol.  We extend this result to a more general class of attacks where the state is arbitrary and the measurement devices have no memory.  We accomplish this by a reduction of arbitrary adversary strategies to qubit strategies and a proof of security for qubit strategies based on the previous proof by Pironio et al. and techniques adapted from Renner.

\end{abstract}

\section{Introduction}
Traditional quantum key distribution protocols rely on a model of the physical devices being used which involves a number of assumptions, such as the dimension of the Hilbert space, the measurement performed, uniform behaviour of detectors, etc..  The actual devices used may deviate from the model and an in-depth knowledge of the system in question is necessary to decide if the assumptions are valid, or the extent to which they are invalid.   If the assumptions are not satisfied, then there exists the possibility of information leaking to the adversary.  Device independent quantum key distribution aims to reduce the number of security assumptions that need to be made in order to obtain a provably secure key from a quantum key distribution protocol.  

Device independent quantum key distribution (DIQKD) aims to replace the model of the physical devices with physically testable or enforceable assumptions.  In particular, the protocols test the extent to which the physical devices can violate a Bell inequality and use this to bound the amount of information leaking to the adversary.  Device independence refers to the fact that no knowledge of the internal mechanism of the devices is necessary, and in fact the devices may be provided by the adversary.  The participants in the protocol only need to observe sufficient violation of a Bell inequality to prove security of the protocol.

Rather than starting from traditional quantum key distribution protocols, DIQKD builds on protocols based on causality constraints.  Work on these protocols began with \cite{Barrett:2005:No-Signaling-an}, and an efficient protocol was introduced in \cite{Acin:2006:Efficient-quant}.  The security proofs for these protocols were generalized to the scenario of global attacks by non-signalling adversaries in \cite{masanes-2006} and \cite{masanes-2008}.  These protocols were first considered in the context of quantum adversaries in \cite{Acin:2007:Device-Independ} with a rigourous proof of security against collective attacks appearing in \cite{Pironio:2009:Device-independ}.

The DIQKD protocols considered to date rely on Bell tests to quantify security.  Currently, assumptions need to be made in order to perform a Bell test and derive a security bound.  Previous proofs applied only to collective attacks, which assume that the devices can be used repeatedly and the different trials are all independent and identical.  The current work aims to weaken the assumptions by allowing the trials to be different and correlated.  The remaining restriction is that there is no memory from one trial to the next.

\subsection{The protocol}

The protocol that we use was originally described in \cite{Acin:2006:Efficient-quant} and shown to be secure against collective quantum attacks in \cite{Acin:2007:Device-Independ} and \cite{Pironio:2009:Device-independ}.  Two parties, Alice and Bob, share a small amount of secret key and wish to expand this into a larger key.  They have access to an uncharacterized device which emits bipartite states, connected by quantum channels to a pair of uncharacterized measurement devices.  Alice's measurement device has three settings, while Bob's has two.  Finally, they have access to a insecure classical channel.  They use some secret key to authenticate data sent on the classical channel.

\begin{enumerate}
	\item Before beginning, Alice randomly chooses a list of $m$ trials to be used for parameter estimation which she sends to Bob encrypted, using some private key bits.
	\item For each trial, Alice and Bob request a state from the source.  If the trial is to be used for parameter estimation, Alice and Bob choose their measurement settings uniformly at random from $\{0,1\}$.  Otherwise Alice chooses setting 2 and Bob chooses setting 0.
	\item After all trials are completed, Alice and Bob announce their measurement settings.
	\item Alice randomly flips each measurement outcome and announces whether or not she does so.  Bob flips his outcomes whenever Alice does.
	\item Alice and Bob publicly announce a permutation and reorder their trials according to this permutation.
	\item Alice and Bob estimate $S$ (defined below) from the parameter estimation trials.
	\item Alice and Bob perform error correction on the remaining trials, correcting Alice's outcomes to correspond with Bob's, resulting in the raw key.
	\item Alice and Bob perform privacy amplification on the raw key according to the secure key rate predicted by $S$.
\end{enumerate}

The above protocol could be efficiently implemented using quantum apparatus by a source of qubit pairs in the state $\ket{\phi_{+}} = \frac{1}{\sqrt{2}}\ket{00} + \frac{1}{\sqrt{2}}\ket{11}$, with Alice's measurements given by the operators $X$, $Y$, and $\frac{X + Y}{\sqrt{2}}$.  Bob's measurement operators are $\frac{X + Y}{\sqrt{2}}$ and $\frac{X - Y}{\sqrt{2}}$.  The security comes from the fact that in order to achieve a high value of $S$, the state that Alice and Bob measure must be close to $\ket{\phi_{+}}$ and hence Bob's measurements are uncorrelated with Eve.  The efficiency of the protocol comes from the fact that Alice can align her measurement with Bob's a significant amount of the time and obtain strongly correlated results, so long as she chooses the other measurements often enough to detect any deviation in the state from $\ket{\phi_{+}}$.

Instead of choosing which trials to use for parameter estimation in advance, Alice and Bob may choose their settings independently, saving some key.  This introduces trials which are unusable (when Alice chooses 2 and Bob chooses 1) and unless Bob chooses 0 and 1 uniformly, there will be some parameter estimation settings that occur more than others.  Conceptually it is easier to suppose that the parameter estimation trials are first chosen and then the settings chosen uniformly.

In \cite{Acin:2007:Device-Independ} and \cite{Pironio:2009:Device-independ} the protocol requires that Alice and Bob symmetrize their data by randomly flipping their outcomes according to a random string which is publicly broadcast.  This simplifies analysis by allowing constraints to be placed on the quantum state.  However, the symmetrization procedure need not be done in practice since it does not change the amount of information leaked to an adversary;  Eve may account for the symmetrization in her own analysis after observing the public random string.  Here we will assume the symmetrization has been done.

\subsection{CHSH inequality}
The CHSH inequality, originally derived in \cite{Clauser:1969:Proposed-Experi}, is a Bell inequality utilizing two measurement settings and two measurement outcomes for two parties.  The two parties, Alice and Bob, each randomly apply one of the two measurement operators to a bipartite state $\rho$ and compare outcomes.  The measurement operators are $A_{a}$ and $B_{b}$, where $a,b \in \{0,1\}$ are the measurement settings for Alice and Bob, respectively.  $A_{a}$ and $B_{b}$ are Hermitian operators with eigenvalues 1 and -1.  The CHSH inequality may be expressed as 
 \begin{equation}
 S = \sum_{a,b = 0,1}\tr \left(A_{a} \otimes B_{b} \rho\right) (-1)^{ab} \leq 2
\end{equation}
for local classical strategies, with an upper bound of $2 \sqrt{2}$ for quantum strategies.  Equivalently, we may use uniformly distributed random variables $a,b \in \{0,1\}$ for the measurement settings and random variables $x, y \in \{0,1\}$ for measurement outcomes, and derive the inequality
 \begin{equation}
p =  P\left(x \oplus y = ab \right) \leq 0.75
\end{equation}
for local classical strategies, with an upper bound of $\cos^{2} \frac{\pi}{8} \sim 0.85$ for quantum strategies.  We say that a trial is successful if $x \oplus y = ab$.  The values $p$ and $S$ are related by
 \begin{equation}
 S = 8p  - 4
\end{equation}
Both of these quantities will be useful in this paper.  We will be interested in the maximum value of $S$ or $p$ achievable by a state $\rho$, maximized over all possible measurements.  We denote these values by $S_{max}(\rho)$ and $p_{max}(\rho)$.

\subsection{Security against collective attacks}
As described above, the protocol could be performed using the same devices over and over.  Pironio et al. (\cite{Pironio:2009:Device-independ}) originally considered security against collective attacks, which relies on the assumption that the devices operate identically each time, and have no memory of the previous trials.  For the source this means that state emitted over $n$ trials has the form $\rho^{\otimes n}$.  A physical implementation using devices that are used repeatedly must meet the following assumptions

\begin{itemize}
	\item On each trial the source emits $\rho$
	\item The combined state that the source emits is $\rho^{\otimes n}$
	\item The measurement devices have no memory
\end{itemize}

 Pironio et al. showed that if Alice and Bob estimate the value of the CHSH operator to be $S$ (settings 0 and 1 for Alice), and they estimate their bit error rate to be $q$ (setting 2 for Alice), then they may extract a secret key at the asymptotic rate of
 \begin{equation}
1 - h \left(\frac{1 + \sqrt{(S/2)^{2} - 1}}{2} \right) - h(q).
\end{equation}

\subsection{Main result and overview of proof}

The main result in this paper is to show that the protocol described in \cite{Acin:2007:Device-Independ} is secure with the same asymptotic key rate against a wider class of attacks.  For our proof we suppose that all trials are performed on separate devices that do not communicate with one another.  This may seem more restrictive, but in fact it is a relaxation since the states may be arbitrary rather than product states, and the devices do not have to operate identically.

A physical implementation of this scheme with many devices is clearly impractical.  A practical implementation with single devices used sequentially could be made, with only the following assumption\footnote{Of course, Alice and Bob's devices must not leak information back to Eve.  Additionally, the measurement devices must not communicate with each other in order to ensure that measurement settings are not leaked.  These conditions are met if we assume that Alice and Bob's labs do not leak information, which is a requirement for any scheme to remain secure.  Additionally, Alice and Bob must have sources of randomness that are uncorrelated with Eve.  Again, this is a basic requirement for any scheme.} :

\begin{itemize}
	\item The measurement devices have no memory
\end{itemize}

\noindent  The model that we use in the proof is that the state is divided into many parts, and each trial corresponds to a measurement that operates on only one part.  By assuming the measurement device has no memory and using it sequentially (providing measurement settings and states one at a time, and receiving the outcome before the next state and setting are given) this condition is enforced.

The source may emit any type of state, which may include a complete specification on how the measurement devices are to operate on a particular trial.  There is no restriction on the dimension of the state or on the form of the measurement operators.  Another important consideration is that there are no losses.  That is to say, there is no provision for cases when no outcome is given.  We may deal with this by assigning a random outcome, which simply adds to the noise, or by adding assumptions, such the adversary having no control over the losses once the measurement settings are given.

The proof relies heavily on \cite{Renner:2005:Security-of-Qua}, chapter 6 and \cite{Pironio:2009:Device-independ}.  Two important contributions are made.  The first is to deal with the unknown dimension of the state, since the finite de Finetti theorem used (\cite{Renner:2007:Symmetry-of-lar} and \cite{Renner:2005:Security-of-Qua}) is sensitive to dimension which in our case is unknown and unbounded.  The second modification is in parameter estimation.  The proof of security in \cite{Renner:2005:Security-of-Qua} assumes that the measurement operators are constant and known, while in our case neither of these is true.  Finally, we adapt the security bounds of \cite{Pironio:2009:Device-independ} to work within Renner's security proof, obtaining the final key rate.

We will first prove security in the case where the states on each trial are restricted to a pair of qubits, which will fix the dimension and allow us to apply the finite de Finetti theorem.  Later we prove that this is sufficient.  

\section{Proof of security for qubit strategies}
In this section we restrict our attention to the case where the state source emits a pair of qubits and the devices each measure one of these qubits.  Our proof of security is derived from the one given by Renner in \cite{Renner:2005:Security-of-Qua}.  The main difference is in the parameter estimation.  Central to the argument is the finite quantum de Finetti theorem published in \cite{Renner:2007:Symmetry-of-lar}.

The first set of states that we will concern our self with are states in the symmetric subspace of $\mathcal{H}^{\otimes n}$ along $\ket{\phi}^{\otimes n-r}$, which is the subspace spanned by states which are of the form $\ket{\phi}^{\otimes n-r} \otimes \ket{\phi^{\prime}}$ for any $\ket{\phi^{\prime}}$ on $r$ subsystems, or any state obtained by permuting the subsystems of such a state.  This subspace is important because the states in it are very close to symmetric product states, which are very nice to work with.  We will denote it by $Sym(\mathcal{H}, \ket{\phi}^{\otimes n-r})$.  The finite quantum de Finetti theorem allows us to break symmetric states into a mixture of these near-product states.

\begin{theorem}[Renner \cite{Renner:2007:Symmetry-of-lar} Theorem 4.3.2]
Let $\rho \in \mathcal{H}^{\otimes n+k}$ be a pure, permutationally invariant state and let $0 \leq r \leq n$.  There exists a measure $\nu$ on the normalized pure states of $\mathcal{H}$, and for each normalized pure state $\ket{\phi}$ in $\mathcal{H}$ a pure density operator $\rho_{\phi}$ on $Sym(\mathcal{H}, \ket{\phi}^{\otimes n-r})$ such that
\begin{equation}
\left|\left| \tr_{k}(\rho) - \int \rho_{\phi}\nu(\phi)  \right|\right|_{1} \leq 2 \exp\left(-\frac{k(r+1)}{2(n+k)} + \frac{1}{2}\dim(\mathcal{H}) \ln k \right)
\end{equation}
\end{theorem}

Here $\tr_{k}$ means tracing out any $k$ subsystems.  The general strategy for the security proof will be to use the fact that the 1-norm is non-increasing under quantum operations combined with the triangle inequality to finally put a bound on the distance between the key obtained by applying the protocol to $\rho$ and the ideal key which is uniform and uncorrelated with Eve.

\subsection{Parameter estimation}

At this point we need to develop techniques for estimating the CHSH value of states which are nearly symmetric product states in the sense introduced in the previous section.  This is analogous to Theorem 4.5.2 in \cite{Renner:2005:Security-of-Qua}.  However, in that case the measurement operations on each subsystem are all known and identical.  In our case the measurements are not in our control, and we may have no description of them.  Fortunately this is not a very important issue.  The CHSH value that can be achieved by a particular state is a property of the state itself.  If the measurements used are not optimal, then the observed CHSH value can only be lower than if the measurements are optimal.  Since we are only interested in lower bounding the CHSH value, this is sufficient.  Any CHSH value that we observe will (leaving statistical fluctuations aside) be a lower bound on the maximum CHSH value achievable by the state.

\begin{lemma}[Parameter estimation]
Let $\ket{\psi} \in Sym(\mathcal{H}_{2} \otimes \mathcal{H}_{2}, \ket{\phi}^{\otimes n + m-r})$ and let $p = p_{max}(\ket{\phi})$ be the maximum expected value for success on the CHSH test on $\ket{\phi}$, optimized over all measurements.  Let $Y$ be the number of successes after conducting the CHSH test on the first $m$ subsystems of $\ket{\psi}$ according to any measurement strategy.  Then for $\mu > 0$

\begin{equation}
P\left(Y/m > p + \mu \right) \leq  \exp{\frac{-2(m\mu - r(1- p))^{2}}{(n-r)cos^{4}\pi/8 } + (n+m)h(\frac{r}{n+m})\ln 2}
\end{equation}

\end{lemma}

The proof has two main steps.  The first is to bound the given probability for states of the form $\ket{\phi}^{\otimes m-r} \otimes \ket{\phi^{\prime}}$, up to permutations of subsystems.  Next we use a lemma of Renner that says $\ket{\psi}$ can be expressed as a superposition of a small number of such states and use another lemma of Renner which bounds how much the probability can change for such superpositions.

\begin{proof}

We now suppose our system is in the state $\ket{\psi^{\prime}}=\ket{\phi}^{\otimes m-r} \otimes \ket{\phi^{\prime}}$ for some $\ket{\phi^{\prime}}$ on $r$ subsystems.  (We may also permute the subsystems without changing the argument.)  Let $X_{j}$ be the random variable corresponding to the success or failure of the CHSH test on the $j$th subsystem for the measurement strategy actually used (which may vary with $j$).  Since the measurement strategy cannot do better than the optimal strategy, we have $ E(X_{j}) < p$ for $1 \leq j \leq m-r$ and $E(X_{j}) < \cos^{2}\frac{\pi}{8} $ for $j > m-r$.  Applying Hoeffding's inequality (\cite{Wikipedia::Hoeffdings-Ineq}) to the first $m-r$ subsystems, we obtain for $t > 1$
 \begin{equation}
 Pr \left( \sum_{j=1}^{m-r}X_{j} > (m-r)(p + t)\right) \leq e^{\frac{-2(m-r)t^{2}}{\cos^{4}\frac{\pi}{8}}}.
\end{equation}
The remaining $r$ subsystems cannot add very much if $r$ is small.  Thus
 \begin{equation}
 Pr \left( \sum_{j=1}^{m}X_{j} > m(p + t) + r\left(1 - p - t\right)\right) \leq e^{\frac{-2(m-r)t^{2}}{\cos^{4}\frac{\pi}{8}}}.
\end{equation}
where $m(p+t) + r (1 - p - t) = (m-r)(p+t) + r$ and the additional $r$ upperbounds the value of $\sum_{j=m-r + 1}^{m} X_{j}$.

We now turn our attention back to $\ket{\psi}$.  Let $z$ be an $m$-tuple with $z_{j} = 1$ if the $j$th trial is successful and $z_{j} = 0$ if it is a failure.  We may write the measurement operator for the CHSH tests together as one large projective measurement $\{M_{z}\}$ with $M_{z}$ the projector corresponding to the outcomes of success and failure given according to $z$.  Then the probability of getting the success/failure outcomes according to $z$ is $\bra{\psi}M_{z}\ket{\psi}$.  Note that $M_{z}$ is positive semi-definite.

We are only interested in the number of successful outcomes, which is given by $w(z)$, the Hamming weight of $z$.  We can restate the above result as 
\begin{equation}
\sum_{w(z) > m(p + t) + r\left(1 - p - t\right)}\bra{\psi^{\prime}}M_{z}\ket{\psi^{\prime}}  \leq e^{\frac{-2(m-r)t^{2}}{\cos^{4}\frac{\pi}{8}}}
\end{equation}

Now suppose that $\ket{\psi}$ is in $Sym(\mathcal{H}, \ket{\phi}^{\otimes n +m- r})$.  We can express $\ket{\psi}$ as a superposition of states of the form $\ket{\phi}^{n+m-r} \otimes \ket{\phi^{\prime}}$ up to permutations of subsystems.  We can apply the above argument to each of these terms in the superposition.  We are only measuring $m$ of the subsystems, so depending on the permutation anywhere between $m-r$ and $m$ of the subsystems may be in the state $\ket{\phi}$.  Note that our bound still applies since the last $r$ subsystems are arbitrary.  The following two lemmas from \cite{Renner:2005:Security-of-Qua} bound how much error may be introduced by this procedure.

\begin{lemma}[Renner \cite{Renner:2005:Security-of-Qua} Lemma 4.5.1]\label{probsuperposition}
Let $\ket{\psi} = \sum_{x \in X} \ket{x}$ and let $P$ be a positive semi-definite operator, then
 \begin{equation}
\bra{\psi} P \ket{\psi}  \leq |X| \sum_{x \in X} \bra{x} P \ket{x}
\end{equation}
\end{lemma}

\begin{lemma}[Renner \cite{Renner:2005:Security-of-Qua} Lemma 4.1.6] \label{symstate}
Let $\ket{\psi}$ be a state in $Sym(\mathcal{H}, \ket{\phi}^{\otimes n - r})$.  Then there exist orthogonal vectors $\ket{x}$, which are permutations of $\ket{\phi}^{\otimes n -r} \otimes \ket{\phi_{x}}$ for $x \in X$ such that $\ket{\psi}$ is in the span of the $\ket{x}$ for various $x$, and $|X| \leq 2^{nh(r/n)}$ where $h(\cdot)$ is the binary Shannon entropy. 
\end{lemma}
 
Applying these results we obtain
\begin{equation}
\sum_{w(z) > m(p + t) + r\left(1 - p - t\right)}\bra{\psi}M_{z}\ket{\psi}  \leq e^{\frac{-2(m-r)t^{2}}{\cos^{4}\frac{\pi}{8}}}2^{(n+m)h(\frac{r}{n+m})}.
\end{equation} 
Rewriting as a probability, we get
 \begin{equation}
 P\left(Y > m(p + t) + r\left(1 - p - t\right) \right) \leq e^{\frac{-2(m-r)t^{2}}{\cos^{4}\frac{\pi}{8}}}2^{(n+m)h(\frac{r}{n+m})}
\end{equation}
or, equivalently
 \begin{equation}
 P\left(Y/m > p + \mu \right) \leq \exp\left({\frac{-2(m\mu - r(1- p))^{2}}{(m-r)cos^{4}\pi/8 } + (n+m)h(\frac{r}{n+m})\ln 2}\right)
\end{equation}

\end{proof}

\subsection{Security}

Security for qubit strategies follows from the same proof as Theorem 6.5.1 in \cite{Renner:2005:Security-of-Qua}, with different parameters.  Since the proof is laid out in great detail in \cite{Renner:2005:Security-of-Qua} we will only sketch the proof and indicate the necessary changes.

We begin with a symmetric state $n + m + k$ pairs of qubits, which we purify (according to Lemma 4.2.2 of \cite{Renner:2005:Security-of-Qua}) on Eve's system to a pure symmetric state $\rho$.  According to the finite quantum de Finetti theorem, we may drop $k$ subsystems and obtain
\begin{equation}
\left|\left| \tr_{k}(\rho) - \int \rho_{\phi}\nu(\phi)  \right|\right|_{1} \leq \frac{2}{9} \epsilon
\end{equation}
with $\rho_{\phi} \in Sym(H_{2}^{\otimes 4}, \ket{\phi}^{\otimes n+m-r})$ and $r$ depending on $n,m,k,\epsilon$ according to table 6.2 of \cite{Renner:2005:Security-of-Qua}.  We next apply parameter estimation by measuring $m$ systems with measurement settings chosen uniformly for Alice and Bob, and determine the number of CHSH successes, $y$.  Then $\frac{y}{m}$ is our estimate of $p$.  If this estimate is below some threshold, $p_{thres} + \mu$ ($p_{thres}$ is used to determine the key rate in the privacy amplification phase) we abort and map the state to 0.  According to Lemma 1, if we choose $\mu$ to be 
 \begin{equation}
 \mu = \frac{4r}{m}\sqrt{\left(-\ln \frac{2 \epsilon}{9} - (n+m) h\left(\frac{r}{n+m} \right) \ln 2 \right)(m-r) \cos^{4}\frac{\pi}{8}}.
\end{equation}
then the true value of $p$ is lower than the estimate minus $\mu$, only with probability less than $\frac{2}{9} \epsilon$.  Thus we may apply the parameter estimation to obtain
\begin{equation}
\left|\left| \rho^{PE} - \int_{V} \rho_{\phi}^{PE}\nu(\phi)  \right|\right|_{1} \leq \frac{4}{9} \epsilon
\end{equation}
where we restrict the integral to the set of states $\ket{\phi}$ which have CHSH probability of success $p_{thres}$ or higher (denoted by $V$).  The $PE$ superscripts indicate the application of the parameter estimation protocol.

We now have (if the protocol did not abort) a state $\rho^{PE}$ which is nearly indistinguishable from a mixture of near-product states with CHSH success probability better than $p_{thres}$.  We may now characterize the smooth min entropy of this family of states and apply privacy amplification, deriving a security bound.  A parameterization of the states appears in \cite{Pironio:2009:Device-independ}, equations (28) through (31).  However, the calculation is essentially the same as it appears in \cite{Renner:2005:Security-of-Qua} and is beyond the scope of this article.  Instead, we will appeal to the final result and calculate the asymptotic key rate.

In \cite{Renner:2005:Security-of-Qua}, Corollary 6.5.2 we find the asymptotic key rate after privacy amplification to be 
 \begin{equation}
 \min_{\sigma_{AB}:S_{max}(\sigma_{AB}) \geq S} H(X|E) - H(X|Y)
\end{equation}
with $H(X|E)$ and $H(X|Y)$ evaluated for state $\sigma_{AB}$, and $S = 8 p_{thres} - 4$, while $X$ and $Y$ are the classical outcomes for Alice and Bob upon measuring $\sigma_{AB}$.  The system $E$ is Eve's system, which we take to be a purification of $\sigma_{AB}$.  Additionally, we must minimize over measurement strategies of Bob's devices consistent with producing a CHSH value of $S$ or better.

We now evaluate the minimum above to obtain the key rate.  First, Lemma 3 in \cite{Pironio:2009:Device-independ} allows us to consider only Bell-diagonal states.  Briefly, the argument relies on the fact that Alice and Bob symmetrize their marginals, together with a suitable local change of basis placing Alice and Bob's measurements on the $X,Z$ plane of the Bloch sphere.  The state $\sigma_{AB}$ can thus be characterized by its eigenvalues, which are the diagonal elements in the Bell basis.  We denote these values by the tuple $\overline{\lambda} = (\lambda_{\Phi_{+}}, \lambda_{\Psi_{-}}, \lambda_{\Phi_{-}}, \lambda_{\Psi_{+}})$, with the subscript denoting the Bell basis element.   Lemma 4 in \cite{Pironio:2009:Device-independ} gives us the bound
 \begin{equation}
h(\overline{\lambda}) - h(\lambda_{\Phi_{+}} + \lambda_{\Phi_{-}}) \leq h \left(\frac{1 + \sqrt{(S_{max}(\sigma_{AB})/2)^{2} - 1}}{2} \right).
\end{equation}
where $h$ is the Shannon entropy.

Recall that for state $\sigma_{XE}$, $H(X|E) = H(\sigma_{XE}) - H(\sigma_{E})$.  The state $\sigma_{E}$ has the same eigenvalues as $\sigma_{AB}$ since $\sigma_{E}$ is the purification.  The eigenvalues are given by $\overline{\lambda}$, so $H(E) = h(\overline{\lambda})$.  

In \cite{Pironio:2009:Device-independ} the state $\sigma_{XE}$ may be calculated from equations (28) through (31) in the proof of Lemma 5.  We sketch the calculation here.  Alice and Bob share the Bell-diagonal state $\sigma_{AB}$ and we give Eve the purification in system $E$ giving a combined state of $\sum_{x} \sqrt{\lambda_{x}} \ket{x}_{AB} \otimes \ket{e_{x}}_{E}$, where $x$ ranges over the Bell states.  We trace out Alice's system and measure Bob's to obtain a classical system $X$ in place of the system $B$.  Bob's measurement can be parameterized on the Bloch sphere as $\cos\phi Z + \sin \phi X$.  The resulting state $\sigma_{XE}$ is specified in equations (30) and (31) in \cite{Pironio:2009:Device-independ} and equation (32) gives the eigenvalues to be\footnote{The states and eigenvalues in \cite{Pironio:2009:Device-independ} are actually for Eve's system conditioned on Bob's measurement outcome, but it is an easy matter to adapt them for our use.}
 \begin{equation}
\Lambda_{\pm} = \frac{1}{4} \left(1 \pm \sqrt{(\lambda_{\phi_{+}} - \lambda_{\psi_{-}})^{2} + (\lambda_{\phi_{-}} - \lambda_{\psi_{+}})^{2} + 2 \cos 2 \phi (\lambda_{\phi_{+}} - \lambda_{\psi_{-}}) (\lambda_{\phi_{-}} - \lambda_{\psi_{+}})} \right)
\end{equation}
each with multiplicity 2.  This gives $H(\sigma_{XE})$ to be $1 + h(\Lambda_{+})$, which is maximized for $\phi = 0$ where $\Lambda_{+} = \lambda_{\phi+} + \lambda_{\phi_{-}}$.  We obtain
 \begin{equation}
 H(X|E) = 1 + h(\lambda_{\Phi_{+}} + \lambda_{\Phi_{-}}) - h(\overline{\lambda}).
\end{equation}
The secret key rate is thus bounded below by
 \begin{equation}
1- h \left(\frac{1 + \sqrt{(S/2)^{2} - 1}}{2} \right) - h(q)
\end{equation}
where $H(X|Y) = h(q)$ and $q$ is the bit error rate between Alice and Bob's raw keys.  This is the same asymptotic rate achieved in \cite{Acin:2007:Device-Independ}.  Note that there is no relationship between $S$ and $q$, since Alice's raw key comes from an unknown measurement.  Her measurement may measure $\rho$ or some other system.  In all cases it is possible for $q$ to range from $0$ to $1$, regardless of the value of $S$.

\section{Security for arbitrary strategies}

\subsection{Block diagonalization of measurement operators}
The following lemma is originally due to Jordan \cite{Jordan:1875:Essai-sur-la-ge}, but has been rediscovered many times.  Modern proofs appear in \cite{Masanes:2006:Asymptotic-Viol} and \cite{Pironio:2009:Device-independ}.  We will use the formulation appearing in \cite{Pironio:2009:Device-independ}.

\begin{lemma}[Pironio et al. \cite{Pironio:2009:Device-independ} Lemma 2]
Let $A^{0}$ and $A^{1}$ be two Hermition operators on $\mathcal{H}$ with eigenvalues 1 and -1.  Then $A^{0}$ and $A^{1}$ can be simultaneously block diagonalized with block sizes $2 \times 2$ and $1 \times 1$.
\end{lemma}

\begin{corollary}\label{qubitobservables}
Let $A^{0}$ and $A^{1}$ be two Hermition operators on $\mathcal{H}$ with dimension $2n$ or $2n - 1$ and eigenvalues 1 and -1, then there exists an isometry $F$ from $\mathcal{H}$ to $\mathcal{H}_{n} \otimes \mathcal{H}_{2}$ and Hermition operators $A^{a,z}$ on $\mathcal{H}_{2}$ with eigenvalues 1 and -1, such that
 \begin{equation}
F(A^{a}) = \sum_{z}\proj{z}{z} \otimes A^{a,z}
\end{equation}

\end{corollary}

This corollary says that we can think of applying one of these two observables as first applying a projection to learn $z$.  The value of $z$ then simultaneously determines a measurement strategy for either measurement setting.  Importantly, the projection onto $z$ can be applied before learning the measurement setting.  This will allow us to consider an arbitrary strategy as a probabilistic combination of qubit strategies.

\subsection{Reduction to qubit strategies}

Let $A_{j}^{a}$ be the observable for Alice mesaurement on the $j$th trial with setting $a$, and analogously for Bob.  We apply corollary \ref{qubitobservables} to pairs of observables  $A_{j}^{0}$ and $A_{j}^{1}$ ($B_{j}^{0}$ and $B_{j}^{1}$) to obtain isometry $F_{j}$ ($G_{j}$), from the Hilbert space of the original state to $\mathcal{Z}^{A}_{j} \otimes \mathcal{H}_{2}$ ($\mathcal{Z}^{B}_{j} \otimes \mathcal{H}_{2}$).  The result is that we can map $A_{j}^{a_{j}}$ ($B_{j}^{b_{j}}$) to 
 \begin{equation}
 \sum_{z_{j}} \Pi_{z_{j}}^{j} \otimes A_{j}^{a,z_{j}}
\end{equation}
with the $\Pi_{z_{j}}^{j}$ commuting for different $j$, and analogously for $B_{j}^{b,w_{j}}$ with projectors $\Pi^{j}_{w_{j}}$.

We have mapped a strategy of Eve to a strategy with state $\rho$ on Hilbert space $\mathcal{Z} \otimes (\mathcal{H}_{2}^{\otimes n})_{A} \otimes (\mathcal{H}_{2}^{\otimes n})_{B}$ with measurement operators of the form above.  Note that we may perform a projective measurement with projectors $\Pi_{z_{j}}^{j}$ for each $j$ to determine all the $z_{j}$ and analogously for Bob's side to determine the $w_{j}$s before determining the measurement setting without changing anything, since these projectors commute with the measurements $A_{j}^{a}$ and $B_{j}^{b}$.  Eve loses nothing by performing this measurement herself, so we may assume that she does so and learns each $z_{j}$ and $w_{j}$.  The result is equivalent to if Eve prepared a mixture of qubit strategies.  We may further suppose that Eve holds the purification for each possible qubit strategy and only increase her power.

We have reduced all possible strategies to a mixture of strategies on qubits.  If Eve in fact performs such a mixture strategy, then for each qubit strategy in the mixture, either the key is secure, or the protocol aborts with high probability.

\section{Open problems}
There are two main open problems left for this protocol of device independent QKD (see \cite{Pironio:2009:Device-independ} for some others).  The first is to remove the restriction that the devices have no memory.  As discussed in \cite{Pironio:2009:Device-independ} the devices may be restricted to classical memory since any quantum memory could instead be teleported forward using extra EPR pairs in the state and classical memory.  

The second open problem is to find an effective means of dealing with channel losses and inefficient detectors.  The detector efficiency loophole quickly translates channel losses and detector inefficiency into low key rates if they are treated as noise.  This severely limits the practicality of the current DIQKD schemes with the present technology.  See \cite{Pironio:2009:Device-independ} for an in-depth discussion.  

\ack
This work is supported by NSERC, Ontario-MRI, OCE, QuantumWorks, MITACS, and the Government of Canada.  Thanks to Antonio Acin and Lluis Masanes for helpful discussions.

\section*{References}

\bibliography{Global_Bibliography}

\newcommand{\etalchar}[1]{$^{#1}$}
\begin{thebibliography}{MRW{\etalchar{+}}06}
 \providecommand{\doi}[1]{{\sc doi}:\href{http://dx.doi.org/#1}{#1}}
 \providecommand{\urlprefix}{{\sc url} }
 \providecommand{\eprintprefix}{{\sc eprint} }

\bibitem[ABG{\etalchar{+}}07]{Acin:2007:Device-Independ}
Antonio Acin, Nicolas Brunner, Nicolas Gisin, Serge Massar, Stefano Pironio,
  and Valerio Scarani.
\newblock Device-independent security of quantum cryptography against
  collective attacks.
\newblock {\em Physical Review Letters}, {\bf 98}(23):230501, 2007.
\newblock \doi{10.1103/PhysRevLett.98.230501}.
\newblock
  \eprintprefix\href{http://arxiv.org/abs/quant-ph/0702152}{arXiv:quant-ph/070%
2152}, \urlprefix\url{http://link.aps.org/abstract/PRL/v98/e230501}.

\bibitem[AMP06]{Acin:2006:Efficient-quant}
Antonio Acin, Serge Massar, and Stefano Pironio.
\newblock Efficient quantum key distribution secure against no-signalling
  eavesdroppers.
\newblock {\em New Journal of Physics}, {\bf 8}(8):126, 2006.
\newblock \doi{10.1088/1367-2630/8/8/126}.
\newblock
  \eprintprefix\href{http://arxiv.org/abs/quant-ph/0605246}{arXiv:quant-ph/060%
5246}, \urlprefix\url{http://stacks.iop.org/1367-2630/8/126}.

\bibitem[BHK05]{Barrett:2005:No-Signaling-an}
Jonathan Barrett, Lucien Hardy, and Adrian Kent.
\newblock No signaling and quantum key distribution.
\newblock {\em Phys. Rev. Lett.}, {\bf 95}(1):010503, Jun 2005.
\newblock \doi{10.1103/PhysRevLett.95.010503}.
\newblock
  \eprintprefix\href{http://arxiv.org/abs/quant-ph/0405101}{arXiv:quant-ph/040%
5101}, \urlprefix\url{http://link.aps.org/doi/10.1103/PhysRevLett.95.010503}.

\bibitem[CHSH69]{Clauser:1969:Proposed-Experi}
John~F. Clauser, Michael~A. Horne, Abner Shimony, and Richard~A. Holt.
\newblock Proposed experiment to test local hidden-variable theories.
\newblock {\em Phys. Rev. Lett.}, {\bf 23}(15):880--884, Oct 1969.
\newblock \doi{10.1103/PhysRevLett.23.880}.
\newblock \urlprefix\url{http://link.aps.org/doi/10.1103/PhysRevLett.23.880}.

\bibitem[Jor75]{Jordan:1875:Essai-sur-la-ge}
Camille Jordan.
\newblock Essai sur la g{\'e}om{\'e}trie {\`a} $n$ dimensions.
\newblock {\em Bulletin de la Soci{\'e}t{\'e} Math{\'e}matique de France}, {\bf
  3}:103, 1875.
\newblock
  \urlprefix\url{http://archive.numdam.org/article/BSMF_1875__3__103_2.pdf}.

\bibitem[Mas06]{Masanes:2006:Asymptotic-Viol}
Lluis Masanes.
\newblock Asymptotic violation of bell inequalities and distillability.
\newblock {\em Physical Review Letters}, {\bf 97}(5):050503, 2006.
\newblock \doi{10.1103/PhysRevLett.97.050503}.
\newblock
  \eprintprefix\href{http://arxiv.org/abs/quant-ph/0512153}{arXiv:quant-ph/051%
2153}, \urlprefix\url{http://link.aps.org/abstract/PRL/v97/e050503}.

\bibitem[Mas08]{masanes-2008}
Lluis Masanes.
\newblock Universally-composable privacy amplification from causality
  constraints, 2008.
\newblock
  \eprintprefix\href{http://arxiv.org/abs/0807.2158v2}{arXiv:0807.2158v2}.

\bibitem[MRW{\etalchar{+}}06]{masanes-2006}
Ll~Masanes, R.~Renner, A.~Winter, J.~Barrett, and M.~Christandl.
\newblock Security of key distribution from causality constraints, 2006.
\newblock
  \eprintprefix\href{http://arxiv.org/abs/quant-ph/0606049v3}{arXiv:quant-ph/0%
606049v3}.

\bibitem[PAB{\etalchar{+}}09]{Pironio:2009:Device-independ}
Stefano Pironio, Antonio Acin, Nicolas Brunner, Nicolas Gisin, Serge Massar,
  and Valerio Scarani.
\newblock Device-independent quantum key distribution secure against collective
  attacks.
\newblock {\em New Journal of Physics}, {\bf 11}(4):045021 (25pp), 2009.
\newblock \doi{10.1088/1367-2630/11/4/045021}.
\newblock \eprintprefix\href{http://arxiv.org/abs/0903.4460}{arXiv:0903.4460},
  \urlprefix\url{http://stacks.iop.org/1367-2630/11/045021}.

\bibitem[Ren05]{Renner:2005:Security-of-Qua}
Renato Renner.
\newblock {\em Security of Quantum Key Distribution}.
\newblock PhD thesis, Swiss Federal Institute of Technology, 2005.
\newblock \eprintprefix\href{http://arxiv.org/abs/quant-ph/0512258}{
  arXiv:quant-ph/0512258v2}.

\bibitem[Ren07]{Renner:2007:Symmetry-of-lar}
Renato Renner.
\newblock Symmetry of large physical systems implies independence of
  subsystems.
\newblock {\em Nature Physics}, {\bf 3}:645 -- 649, July 2007.
\newblock \doi{10.1038/nphys684}.
\newblock
  \eprintprefix\href{http://arxiv.org/abs/quant-ph/0703069}{arXiv:quant-ph/070%
3069},
  \urlprefix\url{http://www.nature.com/nphys/journal/v3/n9/suppinfo/nphys684_S%
1.html}.

\bibitem[Wik]{Wikipedia::Hoeffdings-Ineq}
Wikipedia.
\newblock Hoeffding's inequality.
\newblock \urlprefix\url{http://en.wikipedia.org/wiki/Hoeffding's_inequality}.

\end{thebibliography}
\end{document}